\documentclass[a4paper,twoside,11pt]{article}
\usepackage{amsfonts,amsmath,amssymb,amsthm}
\usepackage{geometry}
\usepackage{booktabs}  

\usepackage[normalem]{ulem}
\usepackage{hyperref}
\hypersetup{
    colorlinks,%
    citecolor=blue,%
    filecolor=black,%
    linkcolor=blue,%
    urlcolor=blue
}
\usepackage{natbib}

\newcommand{\pkg}[1]{{\fontseries{b}\selectfont #1}}
\let\proglang=\textsf
\usepackage{url}

\newtheorem{thm}{Theorem}

\newtheorem{cor}{Corollary}
\usepackage{wrapfig,lipsum,booktabs}

\usepackage{xspace}
\usepackage{bbm}
\usepackage{dsfont}

\def\>{$\Rightarrow$}

\def\R{\mathbb{R}}
\def\L{\mathcal L}
\def\D{D}
\def\W{W}

\def\=>{\Rightarrow}

\def\tr{\mathrm{Tr}}
\def\WT{\mathcal{W}}

\usepackage[table]{xcolor}
\newcolumntype{L}{>{$}l<{$}}
\newcolumntype{C}{>{$}c<{$}}
\definecolor{lgray}{gray}{0.75}
\def\lg{\cellcolor{lgray}}
\definecolor{ggray}{gray}{0.85}
\def\gg{\cellcolor{ggray}}
\definecolor{agray}{gray}{0.95}
\def\ag{\cellcolor{agray}}

\usepackage{graphicx,subfigure}

\RequirePackage{authblk}

\title{Data-driven Thresholding in Denoising\\ with Spectral Graph Wavelet Transform}
\author{
{Basile de Loynes\thanks{Basile de Loynes\\
\hspace*{1.8em}ENSAI, France, 
E-mail: basile.deloynes@ensai.fr
}
, Fabien Navarro\thanks{Fabien Navarro\\
\hspace*{1.8em}CREST, ENSAI, France, 
E-mail: fabien.navarro@ensai.fr}
, Baptiste Olivier\thanks{Baptiste Olivier\\
\hspace*{1.8em}Orange Labs, France. 
E-mail: baptiste.olivier@orange.fr}}
}
\date{\today}

\usepackage{fancyhdr}
\begin{document}
\def\spacingset#1{\renewcommand{\baselinestretch}%
{#1}\small\normalsize} \spacingset{1}
\spacingset{1.45} 

\maketitle
\begin{abstract}
This paper is devoted to adaptive signal denoising in the context of Graph Signal Processing (GSP) using Spectral Graph Wavelet Transform (SGWT). This issue is addressed \emph{via} a data-driven thresholding process in the transformed domain by optimizing the parameters in the sense of the Mean Square Error (MSE) using the Stein's Unbiased Risk Estimator (SURE). The SGWT considered is built upon a partition of unity making the transform semi-orthogonal so that the optimization can be performed in the transformed domain. However, since the SGWT is over-complete, the divergence term in the SURE needs to be computed in the context of correlated noise. Two thresholding strategies called coordinatewise and block thresholding process are investigated. For each of them, the SURE is derived for a whole family of elementary thresholding functions among which the soft threshold and the James-Stein threshold. This multi-scales analysis shows better performance than the most recent methods from the literature. That is illustrated numerically for a series of signals on different graphs. 
\end{abstract}


\section{Introduction}

The emerging field of Graph Signal Processing (GSP) aims to bridge the gap between signal processing and spectral graph theory (see for instance \cite{chung1997spectral,belkin2008towards} and references therein). One objective is to generalize fundamental analysis operations from regular grid signals to irregular structures as graphs. There is an extensive literature on GSP, in particular we refer the reader to \cite{shuman2013emerging} for an introduction to this field and \cite{ortega2018graph} for an overview of recent developments, challenges and applications. As a matter of fact, GSP have already been applied in machine/deep learning: convolutional neural networks (CNN) on graphs \cite{bruna2013spectral,henaff2015deep,defferrard2016convolutional}, semi-supervised classification with graph CNN \cite{kipf2016semi,hamilton2017inductive},  community detection \cite{tremblay2014graph},  to name just a few. In the context of GSP, the authors of \cite{coifman2006diffusion,gavish2010multiscale,hammond2011wavelets} have developed wavelet transforms on graphs. More specifically, in \cite{hammond2011wavelets} a fairly general construction of a frame enjoying the usual properties of standard wavelets is developed: each vector of the frame is localized both in the graph domain and the spectral domain. The transform associated with this frame is named Spectral Graph Wavelet Transform (SGWT). Many studies based on SGWT (or some variants) explore the denoising performance of this approach using different strategies \cite{leonardi2013tight,OnuOnoTan:16,WanShaSmoTib:16,deutsch2016manifold,irion2017efficient,DONG2016561,gobel2018construction} from signal adapted tight frames to regularization method. 

The denoising approach chosen in this paper involves several thresholding processes in the transformed domain of the wavelet coefficients. Actually, this can be seen as an extension to SGWT of the methodology of \cite{donoho1995adapting,Cai:99}. With this approach, the main challenge is the efficient calibration of the parameters minimizing the MSE risk in a complete data-driven way. Recently, in the setting of discrete wavelets transform on a regular grid---the so-called regular case---the Stein's unbiased risk estimate (SURE) has proven to be a powerful tool for signal/image restoration \cite{luisier2007new,pesquet2009sure,vaiter2013local}. Based on the Stein's lemma, this estimator acts as a proxy for the MSE which cannot be computed in practice since the original signal is unknown. In this paper, the SURE is explicitly computed for an arbitrary thresholding process in Theorem~\ref{h-sure} and for correlated noise in the graph domain in Corollary~\ref{cornoise}. Also, let us point out that contrary to the regular wavelet transform, the SGWT is no longer orthogonal so that a white Gaussian noise in the graph domain is transformed in a correlated noise. Consequently, the divergence term of the resulting SURE involves the covariance of the transformed noise making the numerical evaluation less simple than in the regular case. Afterward, the SURE is specified to the case of coordinatewise and block thresholding. The latter is inspired by image denoising problems for which a Stein risk estimator has been proposed in \cite{peyre2011adaptive} to tune both the block-sparsity structure and the threshold. A similar selection strategy has been developed by \cite{navarro2013adaptive} in the context of deconvolution.  The \proglang{R} package \pkg{gasper} which implements the method introduced in this paper is available on github\footnote{https://github.com/fabnavarro/gasper} \citep{gasper} as well as the scripts to reproduce the results presented\footnote{https://github.com/fabnavarro/SGWT-SURE}.

Many denoising methods consist in shrinking the coefficients of a sparse representation of the considered signals in a certain transformed domain. In \cite{TalMil:14}, the aim is to attenuate the coefficients in the Fourier domain corresponding to deletion of the high frequencies contained in the noisy signal. The selection of the parameters is done by minimizing the SURE estimating the MSE in a similar way to our methodology. Due to the limited ability of ideal lowpass graph filter to separate the low-frequency noise, an adaptive weighted graph filter is proposed in \cite{chen2020image}. These two graphs filtering methods can be approximately mimicked with the SWGT based one by removing the finest scales and an appropriate choice of the thresholding. However, the localization properties of SGWT both in space and frequency allow to better process signals with heterogeneous regularity in the initial domain. Intuitively, the thresholding process in the SGWT transformed domain eliminates the high frequency part of the noise where the signal is locally regular and tends to keep the peaks of the original signal whereas the Fourier transform tends to smooth them. This comes at the cost of a redundant frame in which the transformed noise is no longer stationary neither decorrelated.

Regarding the regularization method implemented in \cite{OnuOnoTan:16}, the regularization parameter is also selected optimizing an MSE proxy based on a similar argument. Nonetheless, beyond the fact that the philosophy is different (regularization \emph{versus} thresholding), one stress that the empirical risk bias is explicitly determined while the MSE estimation in \cite{OnuOnoTan:16} is only validated numerically. Another penalization method is given in \cite{WanShaSmoTib:16} that extends the approach from \cite{TibTay:11} within the framework of graphs. For this method, the divergence term is computed explicitly; this gives rise to a data-driven parameter selection method so that this approach is an interesting concurrent to our methodology.  

The paper is organized as follows. Section~\ref{sec:sgwt} introduces the notation and briefly reviews the notions of tight frame and SGWT of \cite{hammond2011wavelets}. Section~\ref{sec:denoise} is devoted to denoising and the SURE estimator for generic thresholding process in the transformed domain. Then, the SURE is specified in the cases of coordinatewise, block thresholding processes and for correlated noise in the graph domain. In Section~\ref{sec:simus} numerical comparisons with the classical Wiener filter (oracle version) and the trend filtering introduced in \cite{WanShaSmoTib:16} for denoising are discussed. Several signals and graphs, including examples from real datasets, are considered. For these experimental results, the construction of the frame follows \cite{gobel2018construction}. In terms of denoising performance, other tight frames such as spectrum adapted and/or signal adapted tight frames from \cite{shuman2015spectrum} and \cite{BehRicvdV:16} might give better results. Still, Theorem \ref{h-sure} actually applies to any tight frame and the question of exhibiting the most efficient one is beyond the scope of the paper (see \cite{shuman2020localized} for a comprehensive survey). 

\section{Spectral Graph Wavelet Transform}
\label{sec:sgwt}
\subsection{Graphs, Frames and Tight Frames}
Let $G$ be an undirected weighted graph, with set of vertices $V$, and weights $(w_{ij})_{i,j\in V}$ satisfying $w_{ij}=w_{ji}$ for $i,j \in V$. The size of the graph is the number of nodes $n=\vert V\vert$. The (unnormalized) graph Laplacian matrix $\L\in\R^{V\times V}$ associated with $G$ is the symmetric matrix defined as $\L=\D - \W$, where $\W$ is the matrix of weights with coefficients $(w_{ij})_{i,j\in V}$, and $\D$ the diagonal matrix with diagonal coefficients $\D_{ii}= \sum_{j\in V} w_{ij}$. A signal $f$ on the graph $G$ is a function $f:V\rightarrow \R$.

Let $\mathfrak F=\{ r_i \}_{i \in I}$ be a frame of vectors of $\mathbb R^V$, that is a family of vectors in $\mathbb R^V$ such that there exist $A,B > 0$ satisfying for all $f \in \mathbb R^V$
\begin{equation} \label{frame-bounds}
A \|f \|^2_2 \leq \sum_{i \in I} |\langle f,r_i \rangle|^2 \leq B \|f\|^2_2.  
\end{equation}
The linear map $T_{\mathfrak F} : \mathbb R^V \rightarrow \mathbb R^I$ defined for $f \in \mathbb R^V$ by $T_{\mathfrak F}f=(\langle f,r_i \rangle)_{i \in I}$ is called the \emph{analysis} operator. The \emph{synthesis} operator is the adjoint of $T_{\mathfrak F}$: namely, it is the linear map $T_{\mathfrak F}^\ast : \mathbb R^I \rightarrow \mathbb R^V$ defined for a vector of coefficients $(c_i)_{i \in I}$ by $T^\ast_{\mathfrak F}(c_i)_{i \in I}=\sum_{i \in I} c_i r_i$. As a frame is in particular a generating family of $\mathbb R^V$, a signal $f \in \mathbb R^V$ can be recovered from its coefficients $T_{\mathfrak F}f$ with the help of the synthesis operator.

\subsection{Construction of Tight Frames}
A frame $\mathfrak F$ is said to be tight if $A=B=1$ in Equation \eqref{frame-bounds}---the latter is then termed the Parseval identity. From now on, the frames considered are supposed to be tight. Let us recall the generic construction of such a frame (\emph{c.f.} \cite{kereta2019monte} for instance).

Since $\L$ is self-adjoint, it admits the spectral decomposition $\L=\sum_{\ell} \lambda_\ell \langle \chi_\ell,\cdot \rangle \chi_\ell,$
where $\lambda_{1} \geq \lambda_{2} \geq \cdots \geq \lambda_{n} = 0$ denote the (ordered) eigenvalues of the matrix $\L$, and $(\chi_{\ell})_{1\leq \ell\leq n}$ are the associated normalized and pairwise orthogonal eigenvectors. Then, for any function $\rho:\mathrm{sp}(\L)\rightarrow \R$ defined on the spectrum $\mathrm{sp}(\L)$ of matrix $\L$, the functional calculus formula reads $\rho(\L) = \sum_{\ell} \rho(\lambda_{\ell}) \langle \chi_{\ell}, \cdot \rangle \chi_{\ell}$.
A finite collection $(\psi_j)_{j=0, \ldots,J}$ is a finite partition of unity on the compact $[0,\lambda_1]$ if $\psi_j : [0,\lambda_1] \rightarrow [0,1]~\textrm{for all}~ j \in \mathcal J~\textrm{and}~\forall \lambda \in [0,\lambda_1],~\sum_{j=0}^J \psi_j(\lambda)=1$.
Given a finite partition of unity $(\psi_j)_{j=0, \ldots, J}$, the Parseval identity implies that the following set of vectors is a tight frame:
\[
\mathfrak F = \left \{ \sqrt{\psi_j}(\L)\delta_i, j=0, \ldots, J, i \in V \right \}.
\]
Also, following \cite{leonardi2013tight,gobel2018construction}, a partition of unity can be easily defined as follows: let $\omega : \mathbb R^+ \rightarrow [0,1]$ be some function with support in $[0,1]$, satisfying $\omega \equiv 1$ on $[0,b^{-1}]$, for some $b>1$, and set
\begin{equation*}
\psi_0(x)=\omega(x)~~\textrm{and}~~\psi_j(x)=\omega(b^{-j}x)-\omega(b^{-j+1}x)~~\textrm{for}~~j=1, \ldots, J,~~\textrm{where}~~J= \left \lfloor \frac{\log \lambda_1}{\log b} \right \rfloor + 2.
\end{equation*}

\subsection{Discrete SGWT Associated with a Partition of Unity}
Let $(\psi_j)_{j=0, \ldots, J}$ be a partition of unity of $[0,\lambda_1]$. The  SGWT of a signal $f \in \mathbb R^V$ is given by
\[
\WT f = \left ( \sqrt{\psi_0}(\L)f^{T},\ldots,\sqrt{\psi_J}(\L)f^{T} \right )^{T} \in \mathbb R^{n(J+1)}.
\]
The adjoint linear transformation $\WT^\ast$ of $\WT$ is:
\[
\WT^\ast \left (\eta_{0}^{T}, \eta_{1}^{T}, \ldots, \eta_{J}^T \right )^{T} = \sum_{j\geq 0} \sqrt{\psi_j}(\L)\eta_{j}.
\]
The tightness of the underlying frame implies that $\WT^\ast \WT=\mathrm{Id}_{\mathbb R^V}$ so that a signal $f \in \mathbb R^V$ can be recovered by applying $\WT^\ast$ to its wavelet coefficients $((\WT f)_i)_{i=1, \ldots, n(J+1)} \in \mathbb R^{n(J+1)}$ (see \cite{hammond2011wavelets}). 

\section{Adaptive Denoising with SGWT}
\label{sec:denoise}
Let $f \in \mathbb R^V$ be some signal on a graph $G$ and $\xi$ be an $n$-dimensional Gaussian vector distributed as $\mathcal N(0,\sigma^2 \mathrm{Id})$. The aim of denoising is to recover the unknown signal $f$ from the observed noisy version $\tilde f = f + \xi$. Basically, our denoising procedure will consist of three steps: (1) compute the SGWT transform $\WT \tilde f \in \mathbb R^{n(J+1)}$; (2) apply a given thresholding operator $h(\cdot)$ to the coefficients $\WT  \tilde f$; (3) apply the inverse SGWT transform to obtain an estimation $\hat f$ of the original signal. Here, the main challenge in denoising consists in choosing a suitable thresholding operator with respect to the noisy signal $\tilde f$ and the underlying graph. The performance measure in the sequel will be the MSE between the original signal $f$ and the denoised signal $\hat f$: $\|f-\hat f\|_2^2$. First, it is worth noting that the Parseval identity allows direct optimization in the transformed domain of wavelet coefficients. Secondly, in practice, obviously the original signal remains unknown. To overcome this difficulty, the MSE is generally substituted with the Stein's Unbiased Risk Estimator which no longer depends on the original signals (see \cite{donoho1995adapting} for instance). Nonetheless, contrary to the usual wavelet transform, the white noise $\xi$ is mapped onto a correlated Gaussian noise. In the next section, the SURE is derived taking into account these correlations.

\subsection{The SURE Estimator in the Transformed Domain}
By linearity, the denoising problem $\widetilde f = f+\xi$ is transferred to the denoising problem $\widetilde F=F+\Xi$ with $\Xi \sim \mathcal N(0,\sigma^2 \WT \WT ^\ast)$, $\widetilde F=\WT \widetilde f$ and $F=\WT f$. It is worth noting that $\WT \WT^\ast$ is an orthogonal projector of rank $n$ in $\mathbb R^{n(J+1)}$ (see \cite{gobel2018construction} for instance). Consequently, its spectral decomposition reads $\WT \WT^\ast=U \Sigma U^\ast$ with $U$ a unitary matrix of $\mathbb R^{n(J+1)}$ and $\Sigma = \left ( \begin{smallmatrix} \mathrm{Id}_{\mathbb R^n} & 0 \\ 0 & 0 \\ \end{smallmatrix} \right )$. 

A thresholding process is a map $h: \mathbb R^{n(J+1)} \rightarrow \mathbb R^{n(J+1)}$. Typically, the map $h$ is a coordinatewise or a block shrinkage in applications. The following result extending the SURE's expression to correlated noise is based on the Stein's lemma in \cite{Ste:81} in which $h$ is assumed to be weakly differentiable. One refers the reader to \cite{Ste:81} for the precise definition.
\begin{thm}[$h$-SURE] \label{h-sure}
  Let $h$ be a weakly differentiable thresholding process for the denoising problem $\widetilde F=F+\Xi$. Then the theoretical MSE is given by
\begin{equation*}
\mathbf E \|h(\widetilde F)-F\|^2 = \mathbf E \left [ -n \sigma^2 + \|h(\widetilde F)-\widetilde F\|^2 + 2 \sum_{i,j=1}^{n(J+1)} \mathbf{Cov}(\Xi_i,\Xi_j) \partial_j h_i(\widetilde F) \right ],
\end{equation*}
where $h_i$ is the $i$-th component of $h$.
\end{thm}
It is worth noting that $\mathbf{Cov}(\Xi_i,\Xi_j)=\sigma^2(\WT \WT ^\ast)_{i,j}$ so that, as soon as the thresholding process $h$ is specified and the noise variance $\sigma^2$ estimated, the SURE of $h$ defined below can be completely computed from the noisy observations as in the regular case:
\begin{equation*}
\mathbf{SURE}(h)=-n \sigma^2 + \|h(\widetilde F)-\widetilde F\|^2+ 2 \sum_{i,j=1}^{n(J+1)} \mathbf{Cov}(\Xi_i,\Xi_j) \partial_j h_i(\widetilde F).
\end{equation*}
\begin{proof}
The theoretical MSE can be rewritten as follows
\begin{equation*}
  \mathbf E \| h(\widetilde F)-F \|^2 = \mathbf E \| h(\widetilde F)-\widetilde F \|^2 + \mathbf E \| \Xi \|^2 + 2\mathbf E \langle h(\widetilde F)-\widetilde F, \Xi \rangle.
\end{equation*}
The second term is equal to $n\sigma^2$ since almost surely $\| \Xi \|^2=\| U^\ast \Xi \|^2=\| P_K U^\ast \Xi \|^2$ where $K= \mathrm{ker}(\WT \WT ^\ast)^\perp$ and $P_K$ the orthogonal projection onto $K$. Finally, setting $g(x)=h(x)-x$, $x \in \mathbb R^{n(J+1)}$, it remains to compute the last term $\mathbf E \langle g(\widetilde F), \Xi \rangle=\mathbf E \langle g(F+\Xi),\Xi \rangle$ where $F$ is deterministic and $\Xi \sim \mathcal N(0,\sigma^2 \WT \WT ^\ast)$. A simple computation gives
\begin{equation*}
\mathbf E \langle g(F+\Xi),\Xi \rangle = \sum_{i=1}^{n(J+1)} \mathbf E[g_i(F+\Xi)\Xi_i]=\sum_{i=1}^{n(J+1)} \mathbf{Cov}(g_i(F+\Xi),\Xi_i).
\end{equation*}
Then, following \cite{Liu:94}, each term in the sum above is given by
\begin{equation*}
  \mathbf{Cov}(g_i(F+\Xi),\Xi_i)=-n \sigma^2 +\sum_{j=1}^{n(J+1)} \mathbf{Cov}(\Xi_i,\Xi_j) \mathbf E[(\partial_j h_i)(F+\Xi)],
\end{equation*}
since $\tr(\sigma^2\WT \WT ^\ast)=n\sigma^2$. This ends the proof of Theorem \ref{h-sure}.
\end{proof}

\subsection{Coordinatewise Thresholding Process}
For a coordinatewise thresholding process, the map $h$ is of the form $h(x)=(\tau(x_i,t_i))_{i=1, \ldots, n(J+1)}$ where $(t_i)_{i=1, \ldots, n(J+1)}$ are the thresholds. In practice, we may choose $\tau(x,t)=x\max \{ 1-t^{\beta}|x|^{-\beta},0 \}$ with $\beta \geq 1$. The most popular choices are the soft thresholding ($\beta=1$), the James-Stein thresholding ($\beta=2$) and the hard thresholding ($\beta=\infty$). The latter will not be considered here since it does not lead to a sufficiently regular thresholding process for Theorem \ref{h-sure} to be applied.

For any $\beta \in  [1,\infty)$, the derivative $\partial_j h_i$ vanishes whereas
\[
  \partial_i h_i(F+\Xi)=\mathbf 1_{[t_i,\infty)}(|\widetilde F_i|) \left [ 1+(\beta-1) \frac{t^\beta}{|\widetilde F_i|^\beta} \right ].
\]
Consequently, the SURE associated with $h$ is given by
\begin{equation} \label{sure-st}
\mathbf{SURE}(h)=-n \sigma^2 + \sum_{i=1}^{n(J+1)} \widetilde F_i^2 \left ( 1 \wedge \frac{t_i^\beta}{|\widetilde F_i|^\beta} \right )^2  + 2 \sum_{i=1}^{n(J+1)} \mathbf{V}(\Xi_i) \mathbf 1_{[t_i,\infty)}(|\widetilde F_i|) \left [ 1+\frac{(\beta-1) t_i^\beta}{|\widetilde F_i|^\beta} \right ].
\end{equation}
The usual expression of the SURE is recovered from the identity above remarking that $\mathbf V[\Xi_i]$ are identically equal to $\sigma^2$ when the transformed noise is uncorrelated.

Let us notice that the coordinatewise soft thresholding ($\beta=1$) satisfies an oracle inequality as shown in \cite{gobel2018construction}. Similarly to the regular case, it states that up to a log factor, the soft thresholding estimator can mimic an oracle projection.

\subsection{Optimization: Donoho and Johnstone's Trick}

The SURE can be optimized in the same way as in the standard case using the Donoho and Johnstone's trick of \cite{donoho1995adapting} whose the justification is recalled below.

For the sake of simplicity, we first consider the case of the coordinatewise thresholding process with a uniform threshold: $t_i=t$ for all $i=1, \ldots, n(J+1)$. Denote by $a_1, \ldots, a_{n(J+1)}$ the absolute values of the noisy wavelet coefficients $|\widetilde F_i|$ in the increasing order. The trick comes from the observation that, on each interval $(a_k,a_{k+1})$, the last term of Equation~\eqref{sure-st} is non-decreasing whereas the second term is an increasing function of $t$. Consequently, the SURE hits its minimum at some value $a_{k^\ast}$, $k^\ast=1, \ldots, n(J+1)$.   

If the thresholds $t_i$ are no longer uniform but merely tied inside blocks with values $t_1, \ldots, t_L$, the same trick is still valid: group the terms in the sums along the different parameters $t_1, \ldots, t_L$ and optimize each partial sum with respect to $t_k$, $k=1, \ldots, L$.

\subsection{Block Thresholding Process}
In order to take advantage of the localization properties of SGWT and the regularity of the original signal, we may introduce block thesholding processes similar to \cite{Cai:99}.

Consider a partition $(B_\ell)_{\ell \in L}$ of $\{ 1, \ldots, n(J+1) \}$ and set $\|x\|_{B_\ell}^2=\sum_{i \in B_\ell} (x_i)^2$. In this case, the thesholding process $h=(h_i)_{i=1, \ldots, n(J+1)}$ reads
\begin{equation*}
h_i(x)=x_i \max \left \{ 1-\frac{t_\ell^\beta}{\|x\|^\beta_{B_\ell}}, 0 \right \}, ~~ x \in \mathbb R^{n(J+1)},~~ \textrm{and} ~~ \ell \in L: i \in B_\ell.
\end{equation*}
If $i,j$ are in different blocks, then $\partial_jh_i$ vanishes. Additionally, if $i,j$ are in $B_\ell$ but $i \neq j$ then
\begin{equation*}
\partial_jh_i(\widetilde F)=\widetilde F_i \mathbf{1}_{[t_\ell,\infty)} ( \|\widetilde F\|_{B_\ell}) \beta t_\ell^\beta \widetilde F_j \|\widetilde F\|_{B_\ell}^{-\beta-2}, 
\end{equation*}
whereas
\begin{equation*}
  \partial_i h_i(\widetilde F)=\mathbf{1}_{[t_\ell,\infty)}(\| \widetilde F \|_{B_\ell})
    \left (1-t_\ell^\beta \|\widetilde F\|^{-\beta}_{B_\ell}+ \beta t_\ell^\beta \widetilde F_i^2 \|\widetilde F\|_{B_\ell}^{-\beta-2} \right ).
\end{equation*}
Consequently, a straightforward computation leads to
\begin{multline*}
  \mathbf{SURE}(h)=-n\sigma^2 + \sum_{\ell \in L} \left ( 1 \wedge \frac{t_\ell^\beta}{\|\widetilde F\|_{B_\ell}^\beta} \right )^2 \|\widetilde F \|_{B_\ell}^2 \\
  + 2 \sum_{\ell \in L} \mathbf 1_{[t_\ell,\infty)}(\|\widetilde F\|_{B_\ell}) \left [ \left (1-\frac{t_\ell^\beta}{\|\widetilde F\|_{B_\ell}^\beta} \right ) \sum_{i \in B_\ell} \mathbf V(\Xi_i)  + \frac{\beta t_\ell^\beta}{\|\widetilde F\|_{B_\ell}^{\beta+2}} \sum_{i,j \in B_\ell} \mathbf{Cov}(\Xi_i,\Xi_j) \widetilde F_i  \widetilde F_j \right ] \\
\end{multline*}
Once again, for uncorrelated transformed noise, the usual expression easily follows from the identity above. Note also that the optimization of the SURE in this case requires more sophisticated techniques as the divergence term is no longer monotone.

\subsection{Correlated Noise in the Graph Domain}

The SURE can also be stated in the context of correlated noise at the cost of some prior information on the covariance structure. More precisely, in the denoising problem $\widetilde f=f+\xi$ with correlated noise, it is supposed that $\xi \sim \mathcal N(0,\Gamma)$ for some covariance matrix $\Gamma$. The denoising problem reads in the transformed problem as $\widetilde F = F + \Xi$ with $\Xi \sim \mathcal N(0,\WT \Gamma \WT^\ast)$.
\begin{cor}\label{cornoise}
  Under the assumption of Theorem \ref{h-sure}, the theoretical $\mathrm{MSE}$ is given by
  \begin{equation*}
    \mathbf E[\|h(\widetilde F)-F\|^2]=\mathbf E \Bigg [-\tr(\WT \Gamma \WT^\ast)
      +\|h(\widetilde F)-\widetilde F\|^2
      +2 \sum_{i,j=1}^{n(J+1)} \mathbf{Cov}(\Xi_i,\Xi_j)\partial_jh_i(\widetilde F) \Bigg ],
  \end{equation*}
  where $h_i$ is the $i$-th component of $h$.
\end{cor}
Let us point out that the parameters selection can be made without computing explicitly $\tr(\WT \Gamma \WT^\ast)$ since it does not depend on $h$---even though, the $\mathrm{MSE}$ estimate is obviously shifted by this quantity. Besides, the correlation structure $\Gamma$ is actually hidden in the quantities $\mathbf{Cov}(\Xi_i,\Xi_j)$, namely, for $1 \leq i,j \leq n(J+1)$: $\mathbf{Cov}(\Xi_i,\Xi_j)=\big ( \WT \Gamma \WT^\ast \big )_{i,j}$.
Consequently, computationally speaking, there is no additional burden compared to the white noise case.
\begin{proof}
The proof follows the lines of Theorem~\ref{h-sure} with $\mathbf E[\Xi_i^2]=\big ( \WT \Gamma \WT^\ast \big )_{i,i},1 \leq i \leq n(J+1).$
\end{proof}
In applications, it is usually reasonable to assume some structure on the covariance matrix $\Gamma$ reflecting the topology of the underlying graph. Typically, the noise on two given vertices may be correlated if those vertices are close enough in the graph. For example, let $\xi_0 \sim \mathcal N(0,\sigma^2\mathrm{Id})$, we set $\xi=\xi_0+\alpha W\xi_0$ where $W$ is the graph matrix of weights and $\alpha > 0$ some tuning parameter describing the global intensity of the correlation. Then, it follows,
\begin{equation} \label{gamma-exemple}
  \Gamma=\sigma^2(\mathrm{Id}+2\alpha W+\alpha^2WW^\ast).
\end{equation}
Other choices are obviously possible.

\subsection{Complexity}
Regarding the space complexity, we need to store the frame and the weights appearing in the SURE for a cost of $O(n^2(J+1)^2)$. With given Laplacian eigendecomposition, the time complexity of the optimization of the SURE is (in average) of order $O(n(J+1)\log(n(J+1)))$ for the coordinate-wise estimator following [8]. For the block estimator, the use of a grid search is a limitation.

\section{Numerical Results}
\label{sec:simus}
This section presents the empirical performance of the proposed automatic threshold selection for different signals defined on different graphs: the Minnesota roads graph (seen as a reference in many recent studies, see \cite{BehRicvdV:16} and references therein) with synthetic signals and the Facebook graph with signals from \cite{WanShaSmoTib:16}, the Pittsburgh Census Tract graph, a graph built from a dataset on New York City taxis with a real signal as well as numerical experiments in the correlated and block cases. All the experiments are conducted with the \proglang{R} package \pkg{gasper}.

\subsection{The Minnesota Roads Graph}
\label{exp-minnesota-1}

The Minnesota roads graph is a planar graph consisting of 2642 vertices and 6606 edges. Each vertex is described by its $(x,y)$-coordinates. The function $\omega$ chosen in the experiments is a piecewise linear function with support in $[0,1]$ and constant equal to 1 on $[0,b^{-1}]$ with $b=2$. From $\lambda_1 \approx 6.89$, we deduce that the number of scales is $J+1=5$---see Section~\ref{sec:sgwt}.

\begin{figure}[h]
\begin{center}
\includegraphics[width=0.45\textwidth]{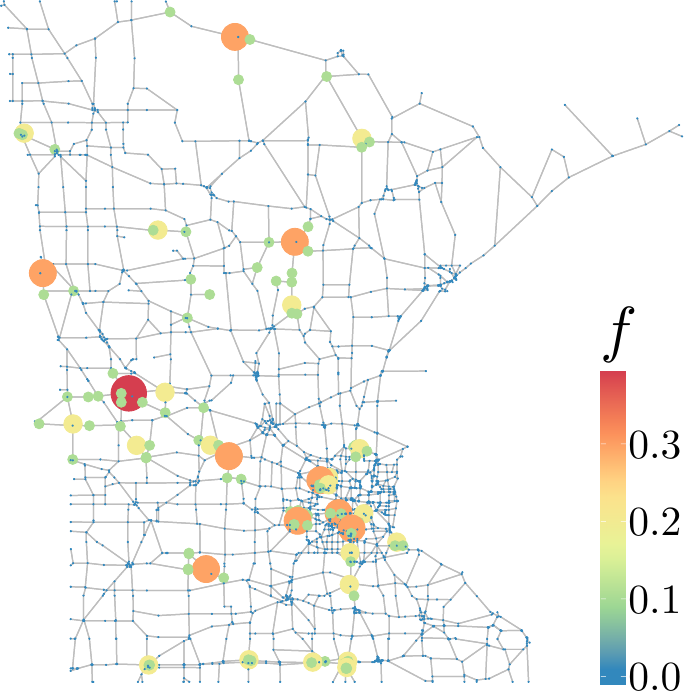}
\includegraphics[width=0.45\textwidth]{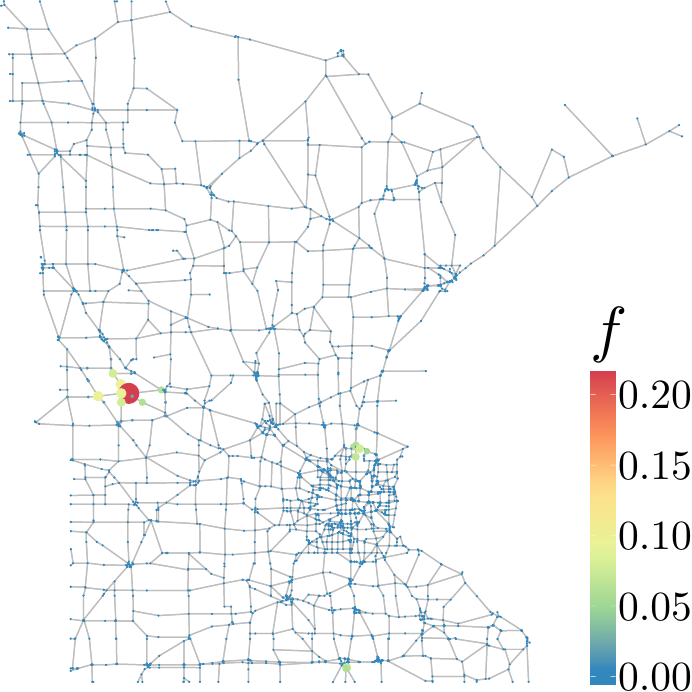}
\end{center}
\caption{Signals used on the Minnesota graph.} 
\label{fig:fig1}
\end{figure}
On this graph, two classes of synthetic signals are generated inspired by the methodology introduced in \cite{BehRicvdV:16}. Let us briefly recall the construction: let $\eta \in (0,1)$ and $k \in \mathbb N$ be two parameters; a signal $f_{\eta,k}$ is obtained by letting the adjacency matrix $W$ acts on an \emph{i.i.d.} realization $x_{\eta}$ of Bernoulli random variables of parameter $\eta$, in symbols $f_{\eta,k}=W^k x_\eta/\lambda_1^k$. This method generates signals with different regularities. This has to be understood in the sense of the graph topology and not that one given by the embedding space $\mathbb R^2$. In the experiment, two signals are generated with parameters $\eta=0.01, k=2$ and $\eta=0.001,k=4$ respectively (see Figure~\ref{fig:fig1} left and right respectively).   

We compare the performance in terms of SNR (computed on the functions after reconstruction) for different denoising strategies, different noise levels and each synthetic signal $f_{\eta,k}$. For each noise level $\sigma=0.005,0.01,0.02$ and $\sigma=0.001,0.002,0.004$, a sample of $N=10$ white Gaussian noise is simulated and a global (G) \emph{versus} level-dependent (LD) coordinatewise thresholding are performed with respect to the soft ($\beta=1$) and James-Stein ($\beta=2$) thresholding rules. For each strategy, we compare the average behavior of the SNR for parameter selected with the oracle ($\mathrm{MSE}^{\beta=1,2}$ obtained by minimizing the MSE using the original signal $f$) and the SURE with known $\sigma$ ($\mathrm{SURE}_\sigma^{\beta=1,2})$. Also, the standard deviation on the sample is provided.


These results are first compared to the classical Wiener filter. More precisely, the Wiener filter consists of attenuating the Fourier coefficient $\mathcal{F}(\widetilde f)$ of $\widetilde f$. Below, we only consider the oracle linear attenuation $\mathcal{F}(\widetilde f)[i]{\mathcal{F}(f)[i]}/({\mathcal{F}(f)[i]^2+\sigma^2})$. While this estimator is unrealistic since it depends on $f$, any Wiener filter has worse performance than this oracle. Table~\ref{tab1} is completed by the performance of the Wiener filter on each signal. Also, the theoretical value $r_{\inf}$ of the oracle risk given in \cite{mallat} is recalled for comparison purpose.

Our methodology is also compared to the so-called graph trend filtering  (\emph{i.e.} for $k=0,1,2$) introduced in \cite{WanShaSmoTib:16}. The graph trend filtering is a regularization method with a penalty term involving the graph difference operator at a given order (see \cite{WanShaSmoTib:16}). In the experiments, we make use of the matlab toolbox \pkg{gtf}\footnote{Available here: https://sites.cs.ucsb.edu/~yuxiangw/resources.html} provided by the authors of \cite{WanShaSmoTib:16}.

\setlength{\tabcolsep}{3.5pt}
 \begin{table*}[h]
 \centering
   \caption{Mean SNR performance over $N=10$ realizations of the low to high noise levels settings with corresponding empirical standard deviation. Left panel: $f_{0.01,2}$ and right panel: $f_{0.001,4}$.}
   \label{tab1}
 \begin{tabular}{rrrr||rrr}
   \toprule
  $\mathrm{SNR}_{\mathrm{in}}$& 16.07$\pm$0.13 & 10.05$\pm$0.13 & 4.03$\pm$0.13 & 16.64$\pm$0.13 & 10.62$\pm$0.13 & 4.60$\pm$0.13\\ 
   \midrule
 $\mathrm{MSE}^{\beta=1,\mathrm{G}}$  
 																	 & 19.04$\pm$0.24 & 14.22$\pm$0.26 & 9.46$\pm$0.26
                                                                & 24.67$\pm$0.33 & 19.77$\pm$0.36 & 14.79$\pm$0.45\\
 $\mathrm{MSE}^{\beta=2,\mathrm{G}}$  
 																    &\gg 20.07$\pm$0.24 &\gg 15.60$\pm$0.30 &\gg 10.69$\pm$0.30
 																	&\gg 26.88$\pm$0.29 &\gg 22.18$\pm$0.37 &\gg 17.08$\pm$0.67\\
 $\mathrm{SURE}^{\beta=1,\mathrm{G}}_{\sigma}$ 
 																					&18.96$\pm$0.27 & 14.16$\pm$0.29 & 9.46$\pm$0.26
 																					 & 24.62$\pm$0.37 & 19.64$\pm$0.48 & 14.70$\pm$0.52\\
 $\mathrm{SURE}^{\beta=2,\mathrm{G}}_{\sigma}$ 
 																					&\ag 20.04$\pm$0.37 & 15.49$\pm$0.43 &\ag 10.64$\pm$0.32
 																					 &\ag 26.73$\pm$0.32 &\ag 21.91$\pm$0.52 &\ag 16.88$\pm$0.59\\
 $\mathrm{MSE}^{\beta=1,\mathrm{LD}}$ 
 																	  & 19.10$\pm$0.24 & 14.28$\pm$0.27 & 9.58$\pm$0.27
 																	  & 24.68$\pm$0.34 & 19.79$\pm$0.36 & 14.83$\pm$0.46\\
 $\mathrm{MSE}^{\beta=2,\mathrm{LD}}$ 
 																	  &\lg 20.08$\pm$0.24 &\lg 15.61$\pm$0.29 &\lg 10.72$\pm$0.30 
 																	  &\lg 26.90$\pm$0.26 & \lg 22.20$\pm$0.36 &\lg 17.13$\pm$0.67\\
 $\mathrm{SURE}^{\beta=1,\mathrm{LD}}_{\sigma}$ 
                																		& 19.10$\pm$0.24 & 14.26$\pm$0.26 & 9.48$\pm$0.24
  																					   & 24.51$\pm$0.40 & 19.59$\pm$0.46 & 14.69$\pm$0.49\\
 $\mathrm{SURE}^{\beta=2,\mathrm{LD}}_{\sigma}$ 
  																						& 20.01$\pm$0.31 &\ag 15.51$\pm$0.36 & 10.61$\pm$0.39
  																						& 26.52$\pm$0.32 & 21.79$\pm$0.34 & 16.73$\pm$0.60 \\
 Wiener & 17.01$\pm$0.13 & 11.87$\pm$0.15 & 7.43$\pm$0.16
            & 17.91$\pm$0.12 & 12.86$\pm$0.13 & 8.40$\pm$0.16 \\ 
  $r_{\mathrm{inf}}$ & 17.05$\pm$0.00 & 11.89$\pm$0.00 & 7.42$\pm$0.00
  								  & 17.96$\pm$0.00 & 12.91$\pm$0.00 & 8.46$\pm$0.00 \\ 
  $\mathrm{MSE}^{k=2}$& 17.35$\pm$0.13 & 11.43$\pm$ 0.16 & 5.68$\pm$0.17
  									    & 19.37$\pm$0.14 & 13.65$\pm$0.18& 8.01$\pm$0.24 \\ 
  $\mathrm{MSE}^{k=1}$& 18.05$\pm$0.14 & 11.98$\pm$0.18 &6.24$\pm$0.18
                                     & 20.43$\pm$0.15 & 14.78$\pm$0.18& 9.30$\pm$0.27 \\ 
  $\mathrm{MSE}^{k=0}$ & 19.57$\pm$0.17 & 13.43$\pm$0.23&7.62$\pm$0.22
                                      &23.38$\pm$0.25 & 17.88$\pm$0.27 &12.80$\pm$0.40 \\ 
    \bottomrule
 \end{tabular}
 \end{table*}

 Generally speaking, we observe from Table~\ref{tab1} that our method performs better than the trend filtering motivating the use of multiscale analysis. This idea is confirmed by the comparison with the Wiener filter, in particular in the lower $\mathrm{SNR}$ regime that is for higher noise levels. Also, similarly to the regular case, numerical experiments shows that the James-Stein threshold ($\beta=2$) is slightly more efficient than the soft threshold in particular for the global thresholding process.

 Let us point out that there is no fundamental difference in terms of performance between the global and level dependent thresholding in this experiment. In fact, the level dependent thresholding always performs at least as good as the global one. Since the additional computational cost is acceptable, the level dependent thresholding appears to be a good choice without any further \emph{a priori} knowledge.

\subsection{The Facebook Graph}
\begin{figure}[!htp]
\begin{center}
\subfigure[Dense Poisson]{
\includegraphics[width=0.35\textwidth]{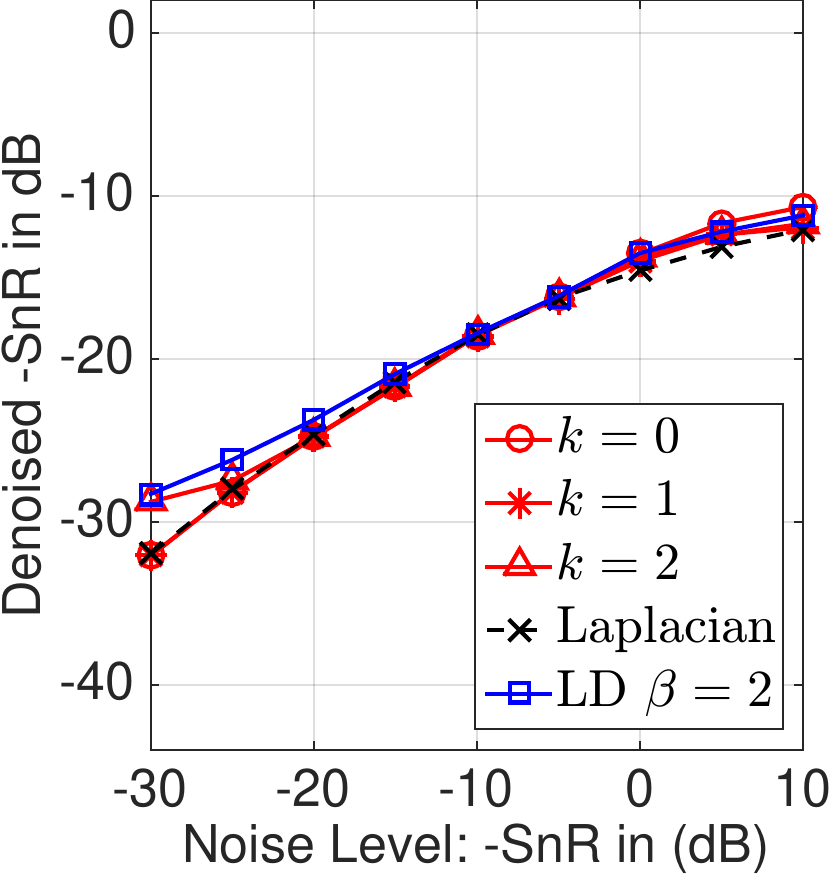}
}
\subfigure[Sparse Poisson]{
\includegraphics[width=0.35\textwidth]{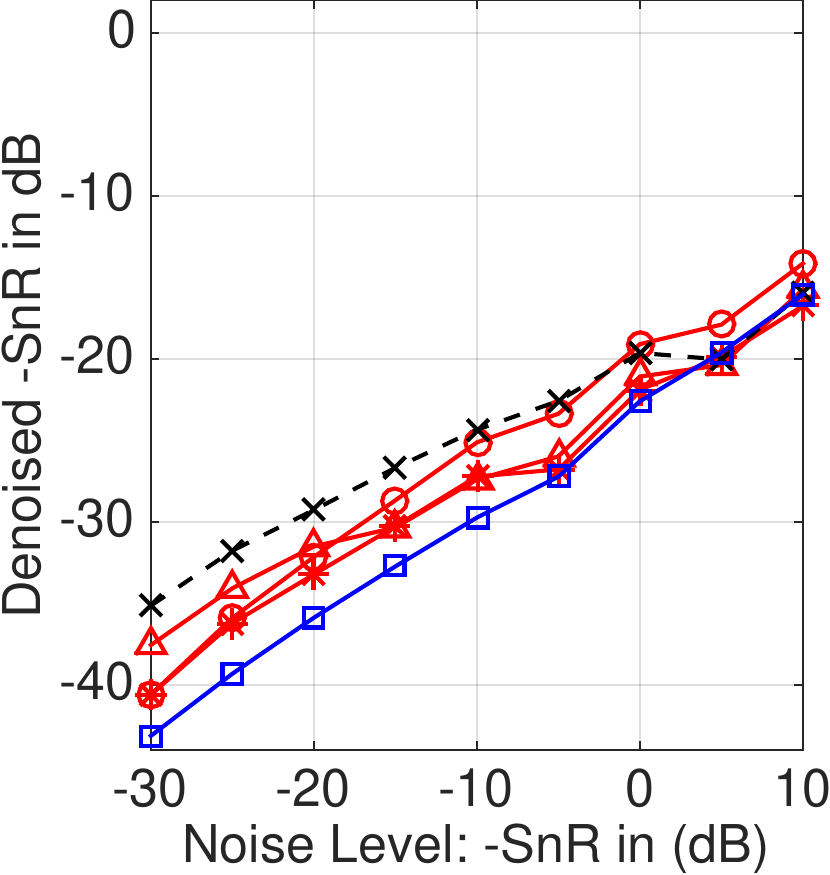}
}
\subfigure[Inhomogeneous]{
\includegraphics[width=0.35\textwidth]{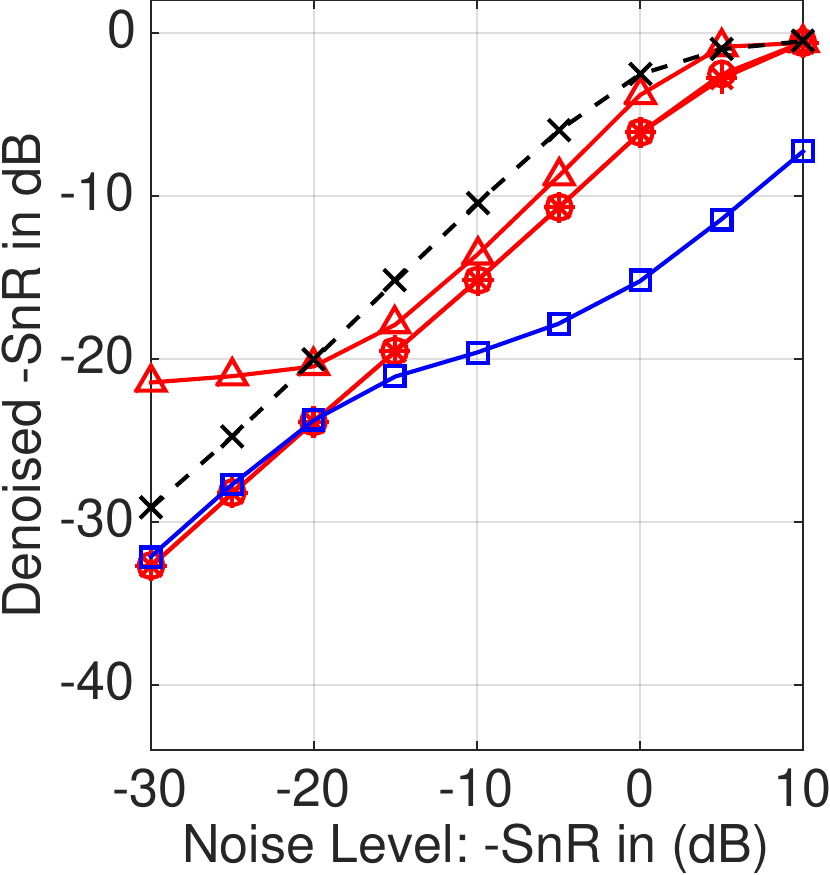}
}
\subfigure[Homogeneous]{
\includegraphics[width=0.35\textwidth]{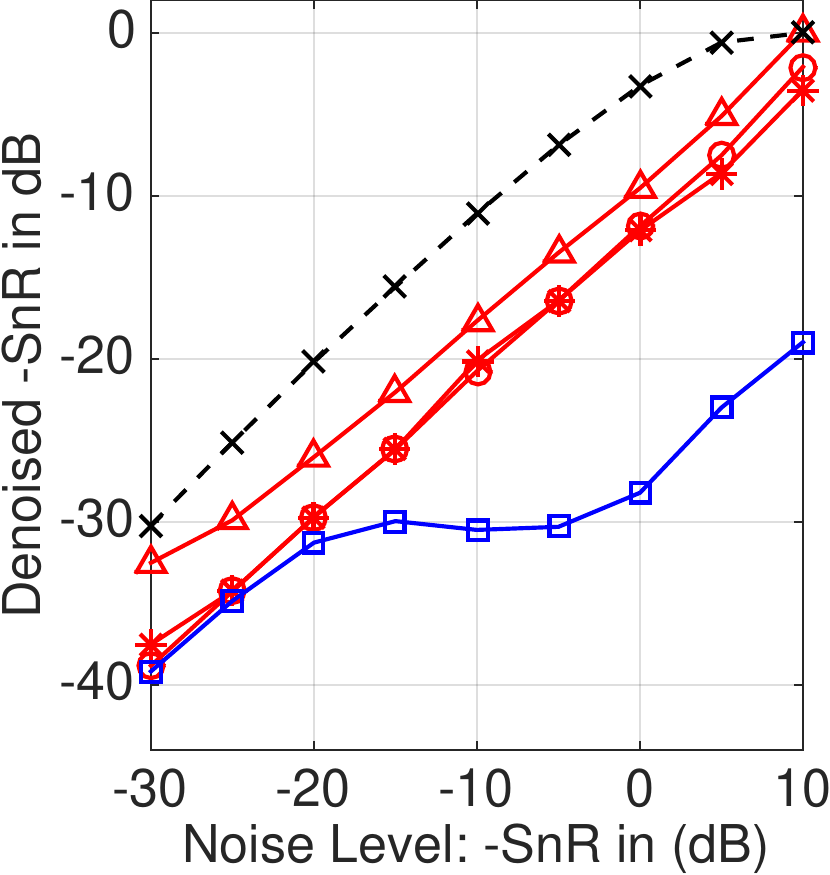}
}
\end{center}
\caption{Mean SNR performance over $N=5$ realizations on the Facebook graph.} 
\label{fig:facebook}
\end{figure}

Here we examined and compare the denoising performance of level dependent SGWT thresholding (LD) against the trend filtering and Laplacian smoothing \cite{smola2003kernels} on a nonplanar graph considered in \cite{WanShaSmoTib:16}: the Facebook graph from the Stanford Network Analysis Project\footnote{\url{http://snap.stanford.edu/data/ego-Facebook.html}}. This undirected graph, collected from survey participants using this Facebook app, is composed of 4039 nodes representing Facebook users, and 88,234 edges representing friendships (see \cite{leskovec2012learning} for more details). For signal $f$, we consider the different regularities used in \cite{WanShaSmoTib:16} as well as the same noise levels, for 5 realizations (see \cite[Section 5.1]{WanShaSmoTib:16} for more details). More precisely, we simply run the scripts provided in the matlab toolbox provided by the authors of \cite{WanShaSmoTib:16}. The results are shown in the Figure~\ref{fig:facebook} (to be compared with \cite[Figure 5 p.14 and Figure 9 p.27]{WanShaSmoTib:16}). 
With the exception of the dense Poisson case, where all methods provide comparable results, LD globally provides better performance than trend filtering (whatever the value of $k$), especially at high noise levels where the maximal gain in terms of SNR is greater than 15dB and exceeds 5dB and 10dB respectively for the 5 highest noise levels for inhomogeneous and homogeneous random walk cases. For this graph, the CPU times associated with trend filtering (for $k=1$ and $k=2$) for one type of signal and 5 realizations (and a 51-point grid search) are of the order of 3 to 4 days (depending on the case), for LD around 25 minutes (including diagonalization and frame calculation which only need to be calculated once).

The calibration of certain parameters has not been studied. However, the latter can considerably influence the performance of the SGWT. For the homogeneous random walk case, we examine here the influence of the number of scales retained for the construction of the frame and controller by $b$. For $b=5,4,3$, the frame contains respectively $7,8,9$ scales. The results are shown in Figure 2, where it can be seen that the frame composed of 7 scales produces the best results. Note that these performances might be improved, for example by making the SURE depend on the $\beta$ parameter characterizing the threshold rule. The frame considered here has only one parameter, other more flexible constructions, based on a partition of the unit or other types of tight frames such as spectrum adapted and/or signal adapted tight frames from \cite{shuman2015spectrum} and \cite{BehRicvdV:16} could also lead to an improvement.

 \begin{figure}[t]
\begin{center}
\includegraphics[width=0.45\textwidth]{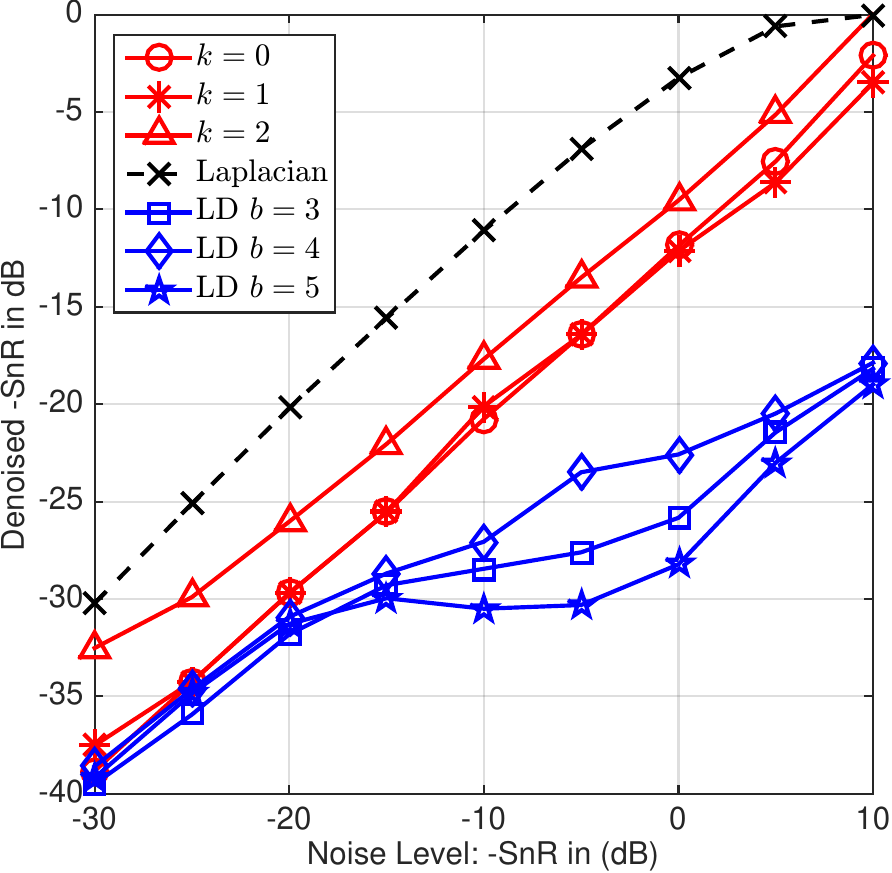}
\end{center}
\caption{Influence of the number of scales on the denoising performance.} 
\label{fig:facebookb}
\end{figure}

\subsection{Pittsburgh Census Tract Graph}
\label{sec:pitt}

For the sake of completeness, our methodology is also compared to the trend filtering on the Pittsburgh Census Tract graph considered in \cite{WanShaSmoTib:16} which consists of 402 vertices and 2382 edges. The very same piecewise linear function $\omega$ with $b=2$ is still used for this graph. The number of scales is then $J+1=7$. For this experiment, only the level dependent threshold procedure is considered.

We consider the signal and 10 realizations of the noisy signal generated in \cite{WanShaSmoTib:16} (corresponding to an average noise level of $4.84\pm0.37\mathrm{dB}$).  The resulting SNR for the oracles of SGWT and fused lasso (\emph{i.e.} $k=0$) are respectively $11.51\pm0.52\mathrm{dB}$ and $9.85\pm0.54\mathrm{dB}$.  

Additionally, we run a comparison with the trend filtering (\emph{i.e.} for $k=0,1,2$)  for the signal $f_{\eta,k}$ with $\eta=0.01$ and $k=5$ with the different noise levels $\sigma=0.004$, $\sigma=0.005$ and $\sigma=0.01$. A comparison with another wavelet estimator proposed in \cite{sharpnack2013detecting} is also provided, considering two thresholding rules (\emph{i.e.} ``soft'' and ``hard''). For these 5 competitors we only report the oracles results.

Even though the SURE no longer depends on the original signal, it does depend on $\sigma^2$ in both methodologies. Since in real applications, the noise level remains unknown in general, we introduce two naive estimators of $\sigma$. In fact, a straightforward computation shows that for any function $g : \mathbb R_+ \rightarrow \mathbb R_+$: 
\begin{equation*}
  \mathbf E[\widetilde f^T g(\L) \widetilde f] = f^T g(\L) f + \mathbf E[\xi^T g(\L) \xi] = f^T g(\L) f + \sigma^2 \tr~g(\L),
\end{equation*}
so that a biased estimator of $\sigma^2$ is given by 
\[
\hat \sigma^2_1 = \frac{\widetilde f^T g(\L) \widetilde f}{\tr~g(\L)}.
\]
As soon as the original signal is reasonably smooth so that $f^T g(\L) f$ is negligible compared to $\tr~g(\L)$, then $\hat \sigma^2$ is an accurate enough estimation of $\sigma^2$. As a first choice, we choose $g(x)=x$. Thanks to Dirichlet's formula, it follows: 
\begin{equation*}
\hat \sigma^2_1 = \frac{\widetilde f^T \L \widetilde f}{\tr~\L}=\frac{\sum_{i,j \in V} w_{ij} |\widetilde f(i)-\widetilde f(j)|^2}{2~\tr~\L}.
\end{equation*}
This is nothing but the graph analogue of the Von Neumann estimator of \cite{vNe:41} explaining the terminology Graph Von Neumann estimator (GVN).

A second natural choice is given by $g(x)=\psi_J(x)$ corresponding to the filter at the finest scale. The resulting estimator is called High Pass Filter Von Neumann (HPFVN). The value of the estimator is easily computed from the coefficients as follows: 
\[
\hat \sigma_2^2=\frac{\sum_{i=nJ+1}^{n(J+1)} (\WT \widetilde f)^2_i}{\tr~\psi_J(\L)}.
\]

 \setlength{\tabcolsep}{4.5pt}
\begin{table}[ht]
\caption{Mean SNR performance over $N=10$ realizations with $f_{0.01,5}$ for the Pittsburgh graph.}
\label{tab:pitt}
\centering
\begin{tabular}{lrrr}
  \toprule
  $\sigma$ & 0.004 & 0.005 & 0.01 \\
  $\hat{\sigma}_1$ &0.0065&   0.0072 & 0.0113\\
  $\hat{\sigma}_2$ & 0.0068&  0.0074 & 0.0114\\
    $\mathrm{SNR}_{\mathrm{in}}$& 9.53$\pm$0.29 & 7.60$\pm$0.29 & 1.58$\pm$0.29\\ 
  \midrule
   $\mathrm{MSE}^{\beta=2,\mathrm{LD}}$&\lg 13.65$\pm$0.29 &\lg 12.22$\pm$0.34 &\lg 8.46$\pm$0.41 \\
   $\mathrm{MSE}^{k=2}$ & 12.28$\pm$0.25 & 11.05$\pm$0.22 & 8.14$\pm$0.29 \\  
   $\mathrm{MSE}^{k=1}$ & 13.10$\pm$0.21 & 11.78$\pm$0.22 & 8.09$\pm$0.36 \\ 
   $\mathrm{MSE}^{k=0}$ & 12.83$\pm$0.23 & 11.42$\pm$0.24 & 7.56$\pm$0.46 \\  
   $\mathrm{MSE}^{\mathrm{Soft}}$ & 11.22$\pm$0.23 & 9.62$\pm$0.27 & 5.22$\pm$0.22 \\  
   $\mathrm{MSE}^{\mathrm{Hard}}$ & 10.22$\pm$0.28 & 8.50$\pm$0.38 & 4.03$\pm$0.14 \\  
   $\mathrm{SURE}^{\beta=2,\mathrm{LD}}_{\sigma}$&13.33$\pm$0.37 & 12.00$\pm$0.30 & 7.95$\pm$0.25 \\ 
   $\mathrm{SURE}^{\beta=2,\mathrm{LD}}_{\hat\sigma_1}$& 12.35$\pm$0.53 & 11.38$\pm$0.62 & 8.13$\pm$0.35\\ 
   $\mathrm{SURE}^{\beta=2,\mathrm{LD}}_{\hat\sigma_2}$& 12.12$\pm$0.57 & 11.27$\pm$0.54 & 8.07$\pm$0.33\\ 
   \bottomrule
\end{tabular}
\end{table}

\begin{figure}[t]
\begin{center}
\subfigure[$\tilde{f}$: $\mathrm{SNR}_{\mathrm{in}}=1.97\mathrm{dB}$]{
\includegraphics[width=0.44\textwidth]{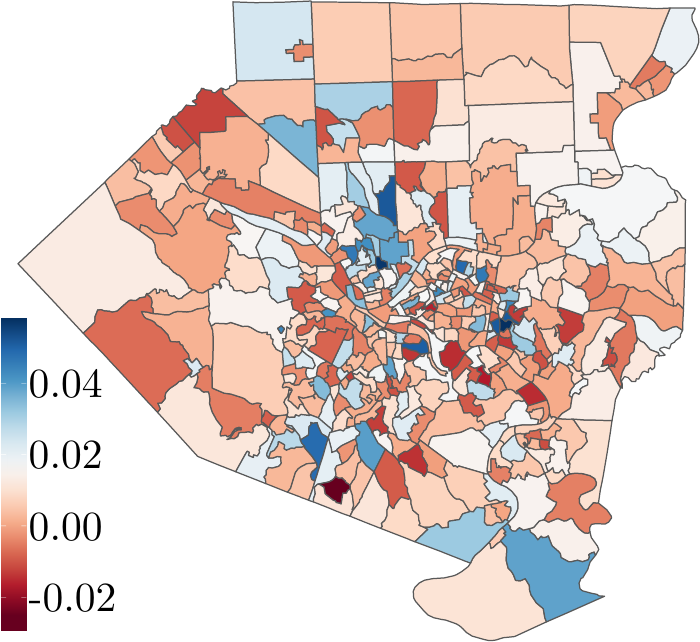}}
\subfigure[$f_{0.01,5}$]{
\includegraphics[width=0.44\textwidth]{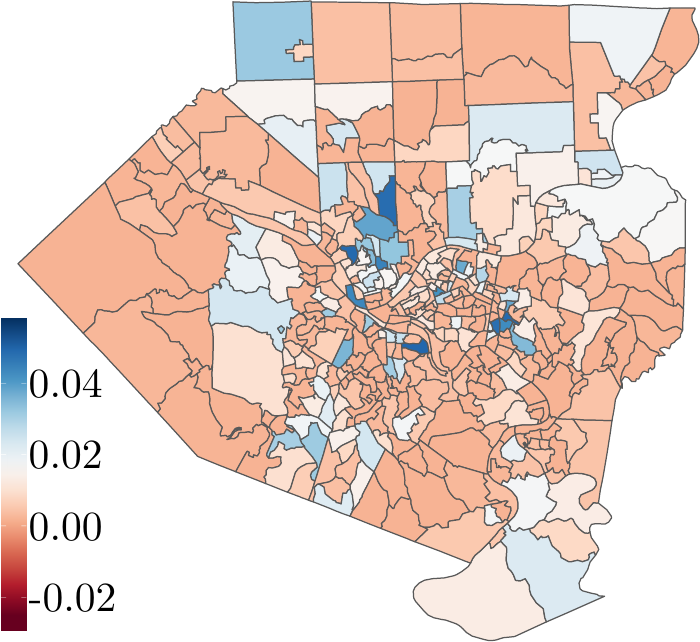}}
\subfigure[$\mathrm{MSE}^{\beta=2}$: $9.75\mathrm{dB}$]{
\includegraphics[width=0.44\textwidth]{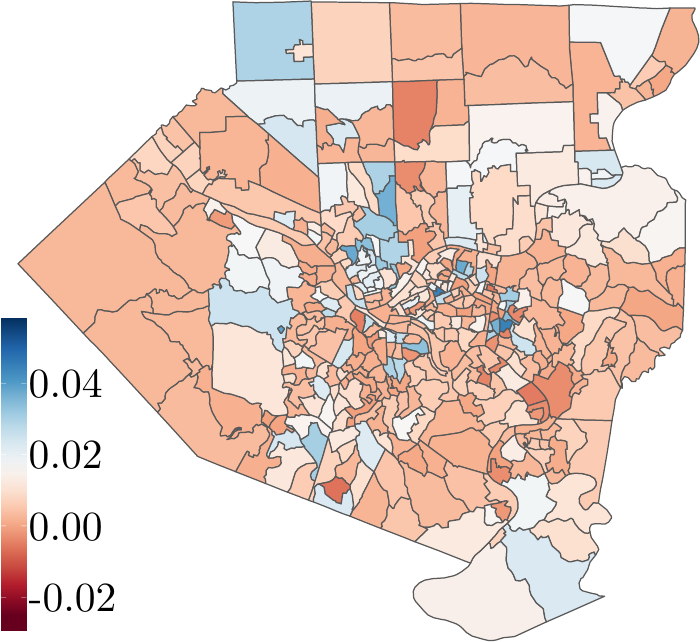}}
\subfigure[$\mathrm{MSE}^{k=0}$: $7.17\mathrm{dB}$]{
\includegraphics[width=0.44\textwidth]{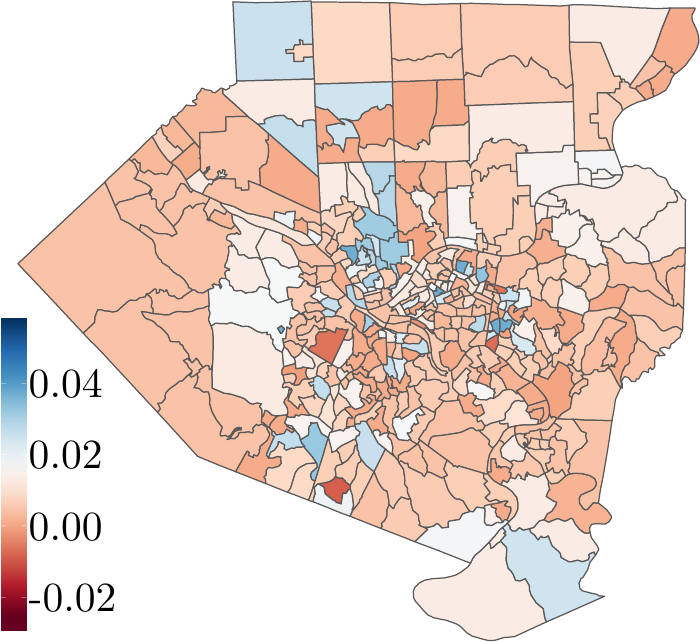}}
\end{center}
\caption{Typical reconstruction for the Allegheny County example.} 
\label{fig:pitt}
\end{figure}

Table~\ref{tab:pitt} summarizes the findings with the nomenclature of Table~1 (where LD stands for level-dependent (LD) coordinatewise thresholding) of the main document. Lines $\mathrm{SURE}_{\hat \sigma_{1,2}}^{\beta=1,2}$ stand for the SURE procedure in which the noise level is estimated by the GVN ($\hat \sigma_1$) and the HPFVN ($\hat \sigma_2$). HPFVN and GVN provide a very comparable sigma estimate in this setting.
Additionally, a visual comparison of our methodology with the fused lasso is illustrated in Figure~\ref{fig:pitt} (corresponding to one realization in the context of the third column of the Table~\ref{tab:pitt}). We can see that our approach provides a gain of about 2.5dB compared to the fused lasso.
 
Again, in these experiments, the multiscale analysis shows better performances than the trend filtering. Besides, the estimation of $\sigma$ is sufficiently accurate to have a fully data-driven procedure.

To conclude, let us stress that the SGWT is computationally more efficient than the GTF ($k=1,2$). For the latter, on a standard laptop (Intel Core i7@2.7GHz-16Go LP-DDR3@2133MHz), each 10 realizations consumed about 4h16m cpu time in mean (min:1h45m, max:6h19m) with the same grid search as for the Pittsburg in \cite{WanShaSmoTib:16}. The CPU consumption for $k=0$ is more decent with a mean of 3m for each 10 realizations exploiting the idea and the c++ code of \cite{chambolle2009total}. Incidentally, this main drawback of GTF was noticed in \cite{padilla2018dfs} forcing a preprocessing of the graph using Depth-first search algorithm. The SGWT on its side consumed 3min for the diagonalization and 42s for each 10 realizations.

\subsection{Real Dataset: New York City Taxis}

Our methodology has been also tested on a real data fetched from NYC taxis\footnote{\url{https://s3.amazonaws.com/nyc-tlc/trip+data/yellow_tripdata_2018-01.csv}} databases. We build a graph with 265 vertices consisting of the LocationID (Pick-Up and Drop-Off) and define Gaussian weights $w_{ij}=\exp(-\tau d_{i,j}^2)$ where $d_{i,j}$ is the mean distance taken on all the trips between $i$ and $j$ or $j$ and $i$. The signal $f$ considered is defined upon the variable ``total amount'' on which an artificial noise is added. For an average input $\mathrm{SNR}_{\mathrm{in}}=5.23\pm0.38\mathrm{dB}$ on $N=25$ observations, we obtain for the level dependent SURE with $\beta=2$ and $\sigma$ known, an output $\mathrm{SNR}$ of $10.41\pm0.62\mathrm{dB}$ compared to the performance of the oracle Wiener filter, $\mathrm{SNR}=7.50\pm0.40\mathrm{dB}$ and the oracle fused lasso, $\mathrm{SNR}=5.69\pm0.43\mathrm{dB}$. 

\subsection{Correlated Noise}

On the Minnesota roads graph and for the signal $f_{\eta,k}$ of Section~4 with $\eta=0.01$ and $k=2$, we add one realization of a correlated noise with covariance matrix given by \eqref{gamma-exemple} where $\alpha=0.5$ is added leading to a noisy signal with $\mathrm{SNR}_{\mathrm{in}}=2.07\mathrm{dB}$. We run the level-dependent coordinate-wise thresholding process. The $\mathrm{SNR}$ given by the oracle involving the unknown signal $f_{\eta,k}$ and the one given by the SURE estimator adapted to correlated noise are similar: $8.54\mathrm{dB}$ \emph{versus} $8.51\mathrm{dB}$. In this case, the SURE for uncorrelated noise shows very bad performances: we found $4.33$ for the corresponding SNR.

The quality of the SURE for uncorrelated noise is closely related to the intensity of the correlation tuned by the parameter $\alpha$. As an example, if we choose $\alpha=0.1$, the SURE adapted to correlated noise still performs very well with an estimated SNR of $10.44\mathrm{dB}$ compared to the oracle $10.46\mathrm{dB}$. The SURE for uncorrelated noise is nonetheless not that bad since it estimates the SNR at $9.78\mathrm{dB}$.

Consequently, the SURE estimate for uncorrelated noise is robust to small correlations which is particularly interesting in applications since it can be difficult to estimate the correlation structure. 

\subsection{Further Experiments with Block Thresholding}

Finally, we report some experimental results in the context of block thresholding in the setting of Section~\ref{sec:pitt}. For each scale $j=0,1,\ldots,7$, the $n$ wavelet coefficients are split into $L$ blocks of uniform length (except for the last block that can be shorter). The best performance of block thresholding is achieved for blocks of size $|L|=47$. The $\mathrm{SNR}_{\mathrm{in}}=9.64\pm0.33\mathrm{dB}$ for 25 realizations. For the oracle global coordinate-wise threshold with $\beta=2$, we obtain a SNR of $11.07\pm0.33\mathrm{dB}$ compared to the block procedure with a uniform threshold $11.82\pm0.40\mathrm{dB}$. The block procedure performs better than the coordinate-wise one for a uniform threshold but is actually worse compared to the level dependent coordinate-wise thresholding process. The level-dependent method is expected to give better performance, but would require a more sophisticated optimization algorithm than grid search to be computationally acceptable.

\section{Conclusion and Perspectives}

In this paper, we have introduced a version of the SURE designed for SGWT, allowing automatic parameter selection in denoising tasks of signals on graphs. Closed-form expressions for coordinatewise and block SUREs have been provided for a wide range of threshold rules. Finally, the case of a correlated noise in the graph has also been considered. Many experiments on the Minnesota graph, the Facebook graph built, the Pittsburgh graph and the NYC taxis graph from real data are conducted. 

For signals of different regularity on those graphs, it has been shown that the SURE provides an efficient estimate of the theoretical MSE. These experiments also show that multi-scale analysis is a serious competitor to existing methods. Indeed, the SGWT shows performances equivalent and sometimes even much better than the GTF, especially at high noise levels.

To be complete, the variance parameter $\sigma^2$ should be estimated. This has been (partially) addressed in the supplementary material by introducing the two estimators HPFVN and GVN. The performance of these estimators highly depends on the underlying signal. Further investigations seem to be necessary. 

Theoretically, the coordinatewise soft thresholding in the transformed domain satisfies an oracle inequality as shown in \cite{gobel2018construction}. Similarly to the regular case, this oracle inequality states that the estimator mimics the oracle projection up to a log factor. The proof relies on the fact that the multivariate risk is expressed as a sum of univariate risks so that the Donoho's machinery applies. Using this fact, a maximal inequality for the SURE might be stated as well.    

Regarding, the numerical complexity, the main limitation is the need of a complete reduction of the Laplacian. In the same vein as \cite{hammond2011wavelets}, many of the involved steps might be numerically optimized using Chebyshev polynomials. Actually, the only problematic step in the method is the computation of the weights $(\WT \WT ^\ast)_{i,j}$ appearing in the SURE. However, their expression in terms of covariance suggests that Monte-Carlo estimation could work. Besides, the space-time complexity might be reduced taking advantage of the low-rank property of $\WT \WT ^\ast$ implying several linear constraints on the weights (precisely $nJ$). Finally, for the block procedure to be completely useful, an adapted optimization algorithm should be implemented. 

Some questions not addressed in the paper remains open. As already announced in introduction of the paper, the choice of a suitable frame for different graphs and different families of signals is still an open problem in spite of advances in recent years. Most likely, good choices of frame should involve a notion of graph limit such as the one introduced for graphons (see \cite{LovLas:12}) or the more probabilistic Benjamin-Schramm limit introduced in \cite{BenSch:01}. This formal study should also give rise to less naïve estimators of the noise level and above all give recommendations according to the class of signals considered.

\bibliographystyle{apalike}
\bibliography{sure_spectral_let_v3}

\end{document}